# High Altitude Platform Station (HAPS)-Aided GNSS for Urban Areas

Hongzhao Zheng, Mohamed Atia, Halim Yanikomeroglu
Department of Systems and Computer Engineering, Carleton University, Ottawa, Canada
hongzhaozheng@cmail.carleton.ca

*Abstract*—Today the global averaged civilian positioning accuracy is still at meter level for all existing Global Navigation Satellite Systems (GNSSs), and the civilian positioning performance is even worse in regions such as the Arctic region and the urban areas. In this work, we examine the positioning performance of the High Altitude Platform Station (HAPS)-aided GPS system in an urban area via both simulation and physical experiment. HAPS can support GNSS in many ways, herein we treat the HAPS as an additional ranging source. From both simulation and experiment results, we can observe that HAPS can improve the horizontal dilution of precision (HDOP) and the 3D positioning accuracy. The simulated positioning performance of the HAPS-aided GPS system is subject to the estimation accuracy of the receiver clock offset. This work also presents the future work and challenges in modelling the pseudorange of HAPS.

## I. INTRODUCTION

The global navigation satellite system (GNSS) has been around for decades. Since the first launch of a legacy GNSS in 1978, the global positioning system (GPS) owned by the US, the positioning accuracy brought by satellites has been improving thanks to the ongoing research in the associated scientific fields. Depending on the application, centimeter level accuracy can be obtained by techniques such as differential GPS (DGPS), real-time kinematic (RTK), multi-constellation GNSS and so forth. For example, the multi-constellation GNSS (BeiDou + Galileo + GLONASS + GPS) has been shown to not only shorten the convergence time, but also to provide centimeter-level positioning accuracy even with 40° cut-off elevation using the precise positioning algorithm [1]. Although numerous techniques have been developed to achieve centimeter-level positioning accuracy, many of which are not suitable for civilian applications such as smartphones, smartwatches, bikes, and so on. Most civilian applications use single frequency and low cost receivers for navigation and positioning, hence precise positioning is not applicable due to reasons such as the incomplete elimination of the ionospheric delay which appears to be one of the largest error sources in the pseudorange measurement. For similar reasons, the most common algorithm used in the civilian applications is therefore the single point positioning algorithm which only requires a single frequency in localization. However the global averaged positioning accuracy of using the single point positioning algorithm is still at meter level. For example, the 95% global averaged horizontal error is less than or equal to 8 m and the 95% global averaged vertical error is less than or equal to 13 m for the GPS system [2]; the 95% global averaged horizontal error is less than or equal to 9 m and the 95 % global averaged vertical error is less than or equal to 10 m for the BeiDou Navigation Satellite System (BDS) [3]; the 95% global averaged horizontal error is less than or equal to 5 m and the 95% global averaged vertical error is less than or equal to 9 m for GLONASS [4]; the 95% global averaged positioning error is less than or equal to 7 m for Galileo [5]. The positioning performance of the GNSS is even worse in the urban area.

Today there are many low Earth orbit (LEO) satellites launched into space, people are also interested in utilizing LEO satellites to aid positioning service. For instance, Li *et al.* prove that the LEO enhanced GNSS can provide centimeter level Signal-In-Space Ranging Error (SISRE) in real-time precise point positioning (PPP) application [6]. Furthermore, researchers have also been investigating the navigation performance which relies exclusively on the LEO satellite signals in case the GNSS signals are unavailable. Khalife *et al.* have shown that a position root mean squared error (RMSE) of 14.8 m for an unmanned aerial vehicle (UAV) can be achieved with only two Orbcomm LEO satellites using the carrier phase differential algorithm [7]. Compare LEO satellites with the typical satellites used in the traditional GNSS which are the medium Earth orbit (MEO) satellites, LEO exhibits the advantages including but not limited to shorter propagation delay and lower pathloss due to shorter distance to the ground user, wider coverage and higher availability due to the enormous number of satellites simultaneously visible/available for positioning. To further enhance high bandwidth networking coverage in challenging areas, High Altitude Platform Stations (HAPS), which resides in the stratosphere with a typical altitude of about 20 km, can be introduced. As urban area is the region where the GNSS positioning performance degrades most severely, we could utilize HAPS as another ranging source by equipping it with a satellite-grade atomic clock on top of a metro city. Since HAPS is only 20 km above the ground, the pathloss of the HAPS signal is expected to be much less than that for any satellites, making the received signal power of HAPS stronger than that of satellite, which likely renders less estimation error in the multipath mitigation of the HAPS signal. HAPS is quasi-stationary as it does not orbit around the globe, this can reduce the number of handovers during the course of positioning. Moreover the HAPS signal does not suffer from the ionospheric effect since it is transmitted from below the ionosphere. Therefore the pseudorange from HAPS can likely be estimated with less error compared with that from satellites. Similar to the pseudorange from satellites which incorporates satellite position error, we should also consider the position error in the pseudorange measurement for HAPS. Fortunately, there are ongoing research in the literature investigating the positioning of HAPS and showing HAPS positioning error can be comparable or even better than the satellite orbit error. For example, a 0.5 m positioning accuracy (circular error probable [CEP] 68 percent) for HAPS has been shown achievable using the modified RTK method [8]. In fact, there are a handful of papers in the literature investigating the HAPS-aided GNSS positioning performance [9]-[12], but none of which considers utilizing HAPS for the sole mission of improving the GNSS positioning performance in the urban area. In this work we examine the HAPS-aided GNSS positioning performance in the urban area via both simulation and physical experiment. For simplicity, the GNSS signal only involves the GPS C/A L1 signal, and the single point positioning algorithm is used to compute the position solution.

## II. SYSTEM MODEL

The general system model is depicted in Fig. 1. The HAPS is situated at 20 km above ground in the stratosphere which is below the ionosphere. There are four satellites shown in Fig. 1, this is just a reminder that at least four satellites are required to perform precise 3D localization using GNSS. Although only a selection of visible satellites is used in position solution calculation in reality, in this work all available satellites are used in position solution calculation for simplicity. The elevation masks for both satellite and HAPS are chosen to be 15 degrees. The pseudorange equation for satellite is given by

$$p_{SAT} = \rho_{SAT} + d_{SAT} + c(dt - dT_{SAT}) + d_{ion,SAT} + d_{trop,SAT} + \epsilon_{mp,SAT} + \epsilon_p \quad (1)$$

where $p_{SAT}$ denotes the satellite pseudorange measurement, $\rho_{SAT}$ is the geometric range between the satellite and receiver, $d_{SAT}$ represents the satellite orbit error, $c$ is the speed of light, $dt$ is the receiver clock offset from GPS time, $dT_{SAT}$ is the satellite clock offset from GPS time, $d_{ion,SAT}$ denotes the ionospheric delay for satellite signals, $d_{trop,SAT}$ denotes the tropospheric delay for satellite signals, $\epsilon_{mp,SAT}$ is the delay caused by the multipath for satellite signals and $\epsilon_p$ is the delay caused by the receiver noise. The pseudorange equation for HAPS is described by

$$p_{HAPS} = \rho_{HAPS} + d_{HAPS} + c(dt - dT_{HAPS}) + d_{trop,HAPS} + \epsilon_{mp,HAPS} + \epsilon_p \quad (2)$$

where $p_{HAPS}$ denotes the HAPS pseudorange measurement, $\rho_{HAPS}$ represents the geometric range between the HAPS and the receiver, $d_{HAPS}$ represents the HAPS position error, $dT_{HAPS}$ is the HAPS clock offset from GPS time, $d_{trop,HAPS}$ denotes the tropospheric delay for HAPS signals, $\epsilon_{mp,HAPS}$ is the delay caused by the multipath for HAPS signals. In this work, the satellite orbit error, the HAPS position error, and the HAPS clock offset are assumed to be zero for simplicity. The simulated vehicle trajectory originates from Carleton University in the suburban area and ends at the Rideau Street of Ottawa in the dense urban area (see Fig. 2). There are four simulated HAPS where one HAPS is following a circular trajectory on top of the downtown Ottawa area, and the other three HAPS are following similar circular trajectories on top of three populated regions near Ottawa. Note that HAPS is quasi-stationary due to factors such as wind, it can move within a confined space. Fig. 3 shows the flowchart of the single point positioning algorithm. Since the HAPS clock offset in this work is assumed zero, we simply use $dT$ to denote the satellite clock offset. From the data collected by the GNSS receiver, we shall obtain both the receiver independent exchange (RINEX) format observation file and the RINEX navigation file, which contains the satellite information such as the pseudorange, the ionospheric parameters, $\alpha$, the Keplerian parameters, and so on. With that information, we know the pseudo-random noise ($PRN$) code which represents the unique number of each satellite, the day of year ($DOY$) which represents the day of year at the time of measurement. Note that $PRN$ is in bold to represent a vector containing the pseudo-random noise code of all visible satellites at the current epoch and the current iteration of estimation. We can compute the satellite positions, $P_{SAT}$, and satellite clock offset, $dT$, using the Keplerian parameters contained in the navigation file. $P_{HAPS}$ denotes a vector containing the positions of all HAPS which are generated using the Skydel GNSS simulator [13], and $p_{HAPS}$ denotes a vector containing the HAPS pseudorange which will be explained in Section III. To compute the position solution, $x$, we firstly initialize the receiver position to the center of the Earth, and the receiver clock offset is initialized to zero. The change in estimate, $dx$, is initialized to infinity. For each epoch of measurement, we will first check if the number of available ranging sources is more than three as at least four ranging sources are required to perform precise 3D localization. Since the receiver position is iteratively estimated, we calculate the elevation angles for both satellites and HAPS with respect to the recently estimated receiver position. Since both the tropospheric delay and the ionospheric delay are functions of the receiver position, these two atmospheric delays are estimated iteratively as well. The elevation angle, satellite pseudorange, HAPS pseudorange, satellite position, satellite clock offset, the tropospheric delay, $d_{trop}$, the ionospheric delay, $d_{ion}$, and the pseudo-random noise ($PRN$) code are modified iteratively based on the re-computed elevation angles for both satellites and HAPS. To prepare the parameters needed for the least square methods, the corrected pseudorange needs to be computed as follows:

$$p^c_{SAT} = p_{SAT} + c \cdot dT - d_{trop,SAT} - d_{ion,SAT} \quad (3)$$

where $p^c_{SAT}$ represents the corrected pseudorange for satellite, $p_{SAT}$ represents the uncorrected pseudorange for satellite. Since the pseudorange error of HAPS is modeled as Gaussian noise representing the estimation residual, the HAPS pseudorange does not need to be corrected. Due to the Earth rotation, the positions of satellites and HAPS at the signal emission time are different from that at the signal reception time, this is known as the Sagnac effect [14]. The coordinates of satellite/HAPS can be transformed from the signal emission time to the signal reception time by [14]

$$\Delta t_{ROT} = t_{rx} - t_{tx} \quad (4)$$

$$P_{i,rx} = M_{ROT}(\omega_E \times \Delta t_{ROT}) P_{i,tx} \quad (5)$$

where $\Delta t_{ROT}$ denotes the signal propagation time, $t_{rx}$ represents the signal reception time, $t_{tx}$ represents the signal emission time, $P_{i,rx}$ is the $i^{th}$ satellite/HAPS coordinates at the signal reception time, $P_{i,tx}$ is the $i^{th}$ satellite/HAPS coordinates at the signal emission time, $\omega_E$ denotes the Earth's rotation rate, and $M_{ROT}(\omega_E \times \Delta t_{ROT})$ is known as the rotation matrix which is described by

$$M_{ROT}(\omega_E \times \Delta t_{ROT}) = \begin{bmatrix} \cos(\omega_E \times \Delta t_{ROT}) & \sin(\omega_E \times \Delta t_{ROT}) & 0 \\ -\sin(\omega_E \times \Delta t_{ROT}) & \cos(\omega_E \times \Delta t_{ROT}) & 0 \\ 0 & 0 & 1 \end{bmatrix}. \quad (6)$$

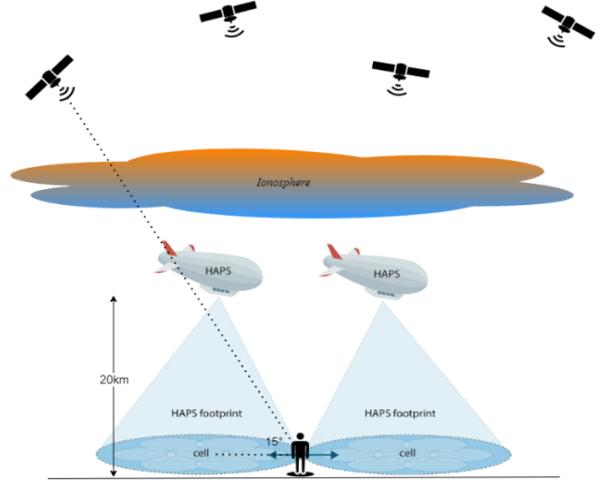

Fig. 1: System model of the HAPS-aided GPS system.

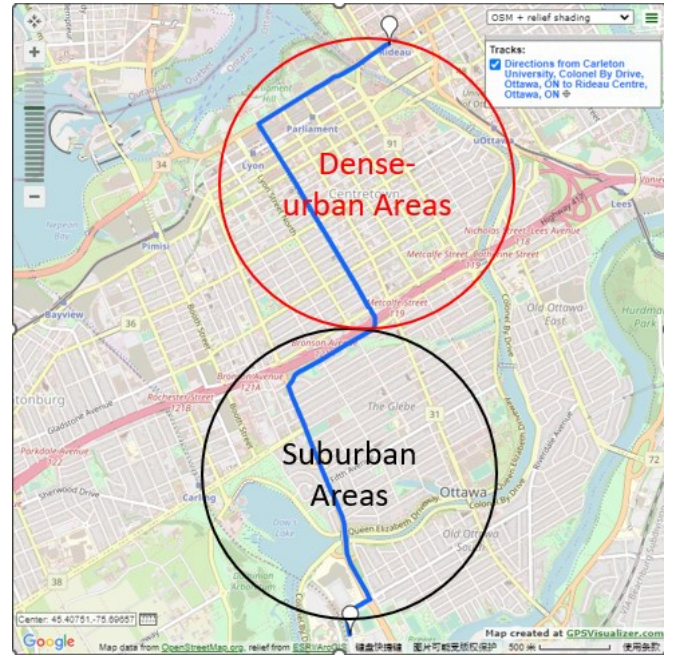

Fig. 2: Vehicle trajectory.

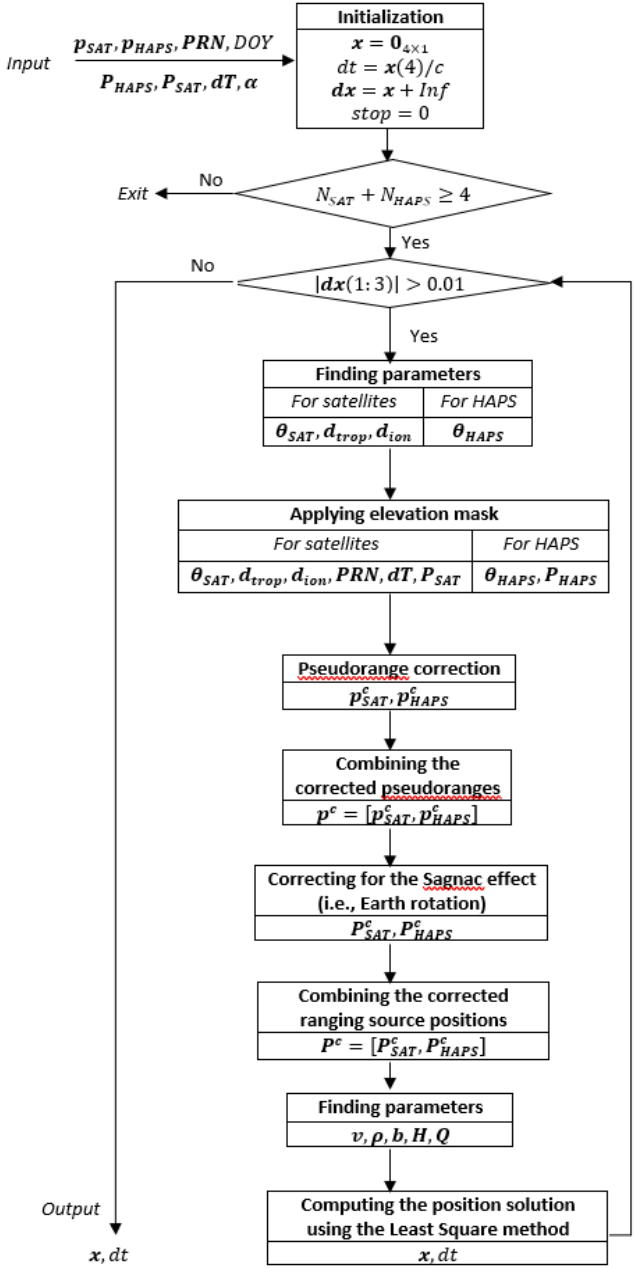

Fig. 3: Flow chart of the single point positioning algorithm.

The line-of-sight vector, $v$, and the geometric range between ranging sources and receiver, $\rho$, are then calculated to compute the a-priori range residual vector $b$ and the design matrix $H$, where

$$b = p^c - \rho \quad (7)$$

$$H = [v, \mathbf{1}_{length(p^c) \times 1}] \quad (8)$$

where $p^c$ is the corrected satellite pseudorange combined with the corrected HAPS pseudorange. At last, the least square solution is computed as

$$Q = (H'H)^{-1} \quad (9)$$

$$dx = QH'b \quad (10)$$

$$dt = dx(4)/c \quad (11)$$

where $Q$ is known as the covariance matrix, $dx(4)$ means the fourth element in the vector $dx$. The covariance matrix, $Q$, is described by

$$Q = \begin{bmatrix} \sigma_x^2 & \sigma_{xy} & \sigma_{xz} & \sigma_{xt} \\ \sigma_{xy} & \sigma_y^2 & \sigma_{yz} & \sigma_{yt} \\ \sigma_{xz} & \sigma_{yz} & \sigma_z^2 & \sigma_{zt} \\ \sigma_{xt} & \sigma_{yt} & \sigma_{zt} & \sigma_t^2 \end{bmatrix} \quad (12)$$

where $\sigma_x$, $\sigma_y$, $\sigma_z$ and $\sigma_t$ represent the standard deviations of the receiver coordinates x, y, z in the Earth-centered Earth-fixed (ECEF) coordinate frame and the receiver clock offset, respectively. The least square solution shall be found when the norm of the change in receiver position, $dx(1:3)$, is sufficiently small. In this work, this threshold is chosen to be 0.01 m. We use the horizontal dilution of precision (HDOP) and the 3D positioning accuracy as the metrics to examine the positioning performance of the proposed HAPS-aided GPS system. To compute the HDOP, we must convert the covariance matrix into the local north-east-down (NED) coordinate frame, which can be done with the following equation [15]:

$$Q_{NED} = R'\tilde{Q}R \quad (13)$$

where $\tilde{Q}$ and $R$ are defined as

$$\tilde{Q} = \begin{bmatrix} \sigma_x^2 & \sigma_{xy} & \sigma_{xz} \\ \sigma_{xy} & \sigma_y^2 & \sigma_{yz} \\ \sigma_{xz} & \sigma_{yz} & \sigma_z^2 \end{bmatrix} \quad (14)$$

$$R = \begin{bmatrix} -\sin \lambda & \cos \lambda & 0 \\ -\cos \lambda \sin \varphi & -\sin \lambda \sin \varphi & \cos \varphi \\ \cos \lambda \cos \varphi & \sin \lambda \cos \varphi & \sin \varphi \end{bmatrix} \quad (15)$$

where $\lambda$ and $\varphi$ represent the longitude and latitude of the receiver, respectively. Then the HDOP is described by

$$HDOP = \sqrt{\sigma_n^2 + \sigma_e^2} \quad (16)$$

where $\sigma_n$, $\sigma_e$, and $\sigma_d$ represent the receiver position errors in the local north, east and down directions, respectively.

## III. SIMULATION OF THE HAPS-AIDED GPS SYSTEM

### A. Simulation Setup

The system model is established using the default Earth orientation parameters of the Skydel GNSS simulation software [13] which considers all GPS satellites orbiting around the Earth and transmitting the L1 C/A code. The Saastamoinen model is chosen to emulate the tropospheric effect and the Klobuchar model is chosen to emulate the ionospheric effect along with the software default Klobuchar parameters (i.e., alpha and beta). The output from Skydel contains the ECEF coordinates of satellites at the signal emission time, the ionospheric corrections, the tropospheric corrections, the satellite clock offsets, the ECEF coordinates of the receiver, the signal emission time, and so forth, at each time stamp from the start of the simulation. The receiver clock offset in the simulation is zero by default. The correction terms in the pseudorange equation of satellite including the satellite orbit error, the multipath and the receiver noise are not separately considered in the simulation, instead a pseudorange error is introduced to reflect the presence of those effect. The pseudorange error of satellite is featured using the built-in first order Gauss-Markov process with the default time constant of 10 s and the standard deviation of 6 m. The continuous model for the first order Gauss-Markov process is described by [16]

$$\dot{x} = -\frac{1}{T_c}x + w \quad (17)$$

where $x$ represents a random process with zero mean, correlation time $T_c$, and noise $w$. The autocorrelation of the first order Gauss-Markov process is described by [17]

$$R(\Delta t) = \sigma^2 e^{-\frac{|\Delta t|}{\tau}} \quad (18)$$

where $\Delta t$ represents the sampling interval, $\sigma$ and $\tau$ denote the standard deviation and the time constant of the first order Gauss-Markov process, respectively. The pseudorange of HAPS is simulated by adding Gaussian noise to the geometric range between HAPS and receiver, where the Gaussian noise represents the sum of all kinds of estimation residuals including the HAPS position, the HAPS clock offset, the tropospheric delay, the multipath and the receiver noise. The pseudorange error for HAPS is modelled using the Gaussian noise with standard deviations of 2 m and 5 m representing the suburban and the dense urban scenario, respectively. The characteristics of the pseudorange errors for the suburban scenario and the dense urban scenario are set to be the same for satellites. Note that by doing this, the positioning performance of the GPS-only system stays the same in both suburban scenario and dense urban scenario. The standard deviation for the HAPS pseudorange error is enforced to be smaller than that for the satellite pseudorange error in both suburban scenario and dense urban scenario. All the available satellites (i.e., satellites with elevation angles greater than the predefined elevation mask) are simultaneously utilized for positioning as if all satellites above the elevation mask are in line of sight (LOS) with the receiver. Under this setting, we examine the 3D positioning performance for the GPS-only system, the one-HAPS with GPS system, the four-HAPS with GPS system and the four-HAPS-only system. For the one-HAPS with GPS system, we use the HAPS on top of the downtown Ottawa area which elevation is above 80°.

### B. Simulation Results

The cumulative distribution functions of the 3D positioning accuracy for different systems with the standard deviations of the HAPS pseudorange error being 2 m and 5 m are shown in Fig. 4 and Fig. 5, respectively. From Fig. 4, we can observe that with much less pseudorange error for HAPS, the four-HAPS with GPS system achieves the best positioning performance, the one-HAPS with GPS system achieves almost the same positioning performance as the GPS-only system, and the four-HAPS-only system achieves slightly worse performance than the four-HAPS with GPS system. The reasons why the four-HAPS-only system does not achieve the best positioning performance is potentially due to the following reasons 1) it has much fewer ranging sources in receiver position computation; 2) the ranging source geometry is poor as the elevation angles for all four HAPS at any given time are above 40° with one even above 80°. From Fig. 5, we see that with the HAPS pseudorange error similar but slightly smaller than the satellites' pseudorange error, the four-HAPS-only system achieves the worst positioning performance but the four- HAPS with GPS

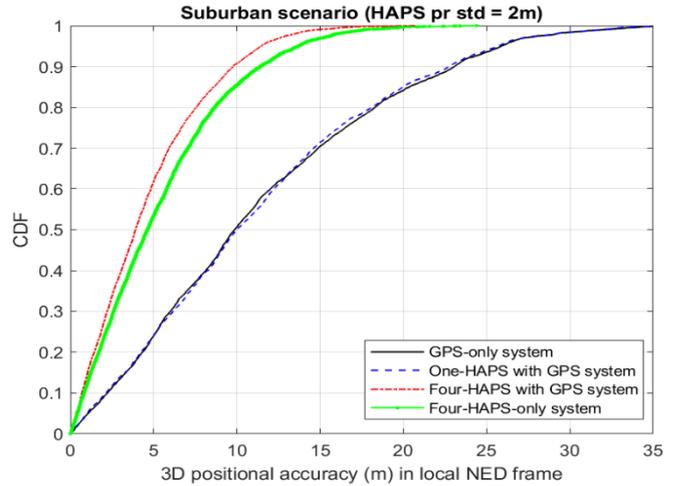

Fig. 4: CDF for 3D position accuracy (suburban scenario).

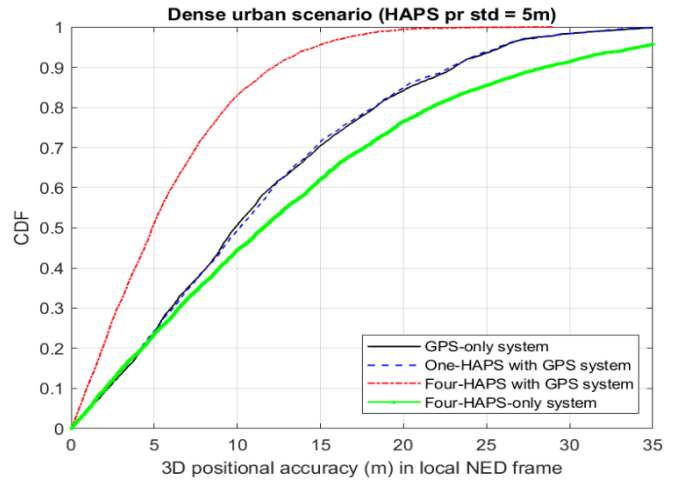

Fig. 5: CDF for 3D position accuracy (dense urban scenario).

system still outperforms the other systems considered.

## IV. FIELD EXPERIMENTS

### A. Experiment Setup

To verify and support the simulation results, we also process the raw GNSS data collected along the vehicle trajectory which is similar to the one shown in Fig. 2 with a slight difference due to partial road closure on the day of data collection. The raw GNSS data are collected using the Ublox EVK-M8T GNSS unit and processed using the single point positioning package developed by Napat Tongkasem [18] with proper modification so that HAPS can be incorporated in the single point positioning algorithm. Table I gives the specifications of the EVK-M8T GNSS unit. To reflect realistic LOS conditions for HAPS, the LOS probability with respect to the HAPS elevation angle in the urban area is implemented based on [19] and [20]. Note that the LOS probability for HAPS provided by [19] is generated based on the city of Chicago and enforcing the LOS probability on HAPS in the dense urban area in Ottawa might be too harsh considering the incompatible city scale. The pseudorange of HAPS in the experiment is modeled as the addition of the geometric range between the satellite and receiver, the receiver clock offset multiplied by the speed of light and the pseudorange error representing the sum of all kinds of estimation residuals. The pseudorange errors for HAPS in the suburban area and in the dense urban area are simulated as Gaussian noise with standard deviations of 2 m and 5 m, respectively. Since the vehicle trajectory involves both suburban area and dense urban area, the entire route is divided into two parts where the first part is considered as the suburban scenario and the second part is considered as the dense urban scenario (see Fig. 2). By observing the positioning performance of the GPS-only system using the real GPS data, the LOS probability for the suburban area is applied to HAPS for epochs less than 380 s, and the LOS probability for the dense urban area is applied to HAPS for epochs greater than or equal to 380 s (refer to Fig. 6). Since the GNSS receiver does not provide accurate receiver clock offset with respect to the GPS time, the receiver clock offset in each epoch is estimated by making use of the ground truth receiver position. The ground truth data is provided by Ublox EVK-M8U GNSS unit, which is equipped with accelerometer and gyroscope, hence it can perform sensor fusion to get better positioning performance and dead reckoning when the signal quality degrades.

TABLE I. EVK-M8T GNSS UNIT SPECIFICATIONS [21]

| Parameter | Specification |
|---|---|
| Serial Interfaces | 1 USB V2.0 |
| | 1 RS232, max.baud rate 921,6 kBd DB9 +/- 12 V level 14 pin – 3.3 V logic |
| | 1 DDC (I2C compatible) max. 400 kHz |
| | 1 SPI-clock signal max. 5,5 MHz – SPI DATA max. 1 Mbit/s |
| Timing Interfaces | 2 Time-pulse outputs |
| | 1 Time-mark input |
| Dimensions | $105 \times 64 \times 26$ mm |
| Power Supply | 5 V via USB or external powered via extra power supply pin 14 (V5_IN) 13 (GND) |
| Normal Operating Temperature | $-40°C$ to $+65°C$ |

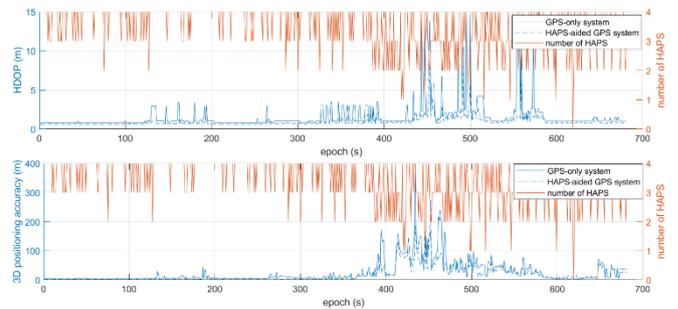

Fig. 6: HDOP (top) and 3D position accuracy (bottom).

### B. Experiment Results

Fig. 6 shows the HDOP, and the 3D positioning accuracy overlapped with the number of visible HAPS at each epoch. As we can see from Fig. 6, the HDOP and 3D positioning accuracy of the HAPS-aided GPS system are better than that of the GPS-only system in both suburban area and dense urban area. Moreover, we can observe that the positioning performance of the HAPS-aided GPS system is more stable than the GPS-only system as there are less spikes for the HAPS-aided GPS system. Note that, the pseudorange of HAPS in the experiment is modeled as a function of the receiver clock offset, which is estimated with the best effort, additional error should be expected in the pseudorange of HAPS with the magnitude depending on the quality of all visible satellite signals and the ground truth receiver position. As we would expect the quality of the satellite signals in the suburban area is better compared to that in the dense urban area, the receiver clock offset would also be expected to be estimated with higher accuracy in the suburban area than in the dense urban area, hence the HDOP of the HAPS-aided GPS system in the suburban area is better. The cumulative distribution functions of the 3D positioning accuracy in the suburban and dense urban areas are shown in Fig. 7 and Fig. 8, respectively. From Fig. 7 and Fig. 8, we can observe that the HAPS-aided GPS system outperforms the GPS-only system, especially in the suburban area.

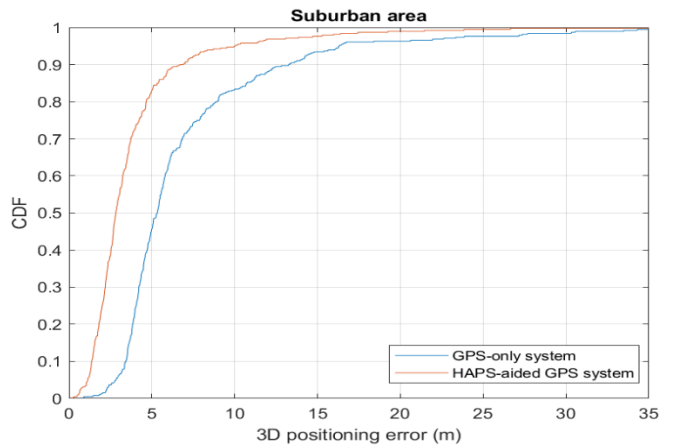

Fig. 7: CDF of 3D position accuracy in the suburban area.

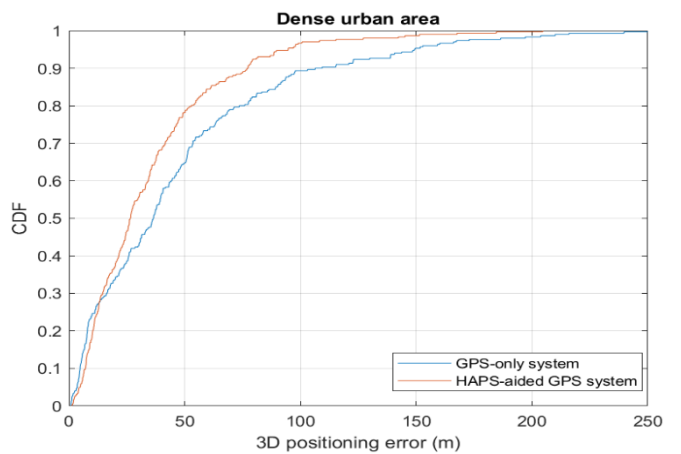

Fig. 8: CDF of 3D position accuracy in the dense urban area.

## V. Conclusion

As we are passing 5G and soon entering 6G and beyond, HAPS can be of invisible treasure as it can be used for computation offloading [22], edge computing [23], even base station [24] to meet human needs. HAPS can be another type of ranging source which is quasi-stationary and much closer to the ground of the Earth. Compared to satellite, HAPS exhibits the advantages of lower latency, lower pathloss, lower pseudorange error, and it can provide continuous coverage to reduce the number of handovers for the users in a certain region. Since urban area is the region where GNSS positioning performance degrades severely and where most people live in, deploying several HAPS acting as another type of ranging source on top of a metro city would improve the GNSS positioning performance and maximize the value of the extra payload on HAPS. The HAPS-aided GNSS can also be deployed in the regions with extreme environment such as the Arctic region where the satellite availability is low, and the ionospheric disturbances is severe [25]. From both the simulation and physical experiment results, we observe that HAPS can indeed improve the 3D positioning accuracy, especially in the suburban area. To improve the results of HAPS-aided GPS system in the dense urban area, the receiver clock offset should be estimated with higher accuracy. In future work, the received signal powers of HAPS and satellite will jointly be considered, a satellite selection algorithm will be applied to better emulate the way a modern GNSS receiver processes the raw GNSS data.

## Acknowledgment

This paper is supported in part by Huawei Canada. The Skydel software is a formal donation from Orolia.

## References

[1] X. Li et al., "Precise positioning with current multi-constellation Global Navigation Satellite Systems: GPS, GLONASS, Galileo and BeiDou," *Sci Rep 5*, 8328 (2015).

[2] "Global Positioning System standard positioning service performance standard," *GPS.GOV*, USA, Apr. 2020. Accessed on: Jul. 9, 2022. [Online]. Available: https://www.gps.gov/technical/ps/2020-SPS-performance-standard.pdf

[3] "BeiDou Navigation Satellite System open service performance standard," *China Satellite Navigation Office*, China, May 2021. Accessed on: Jul. 9, 2022. [Online]. Available: http://en.beidou.gov.cn/SYSTEMS/Officialdocument/202110/P020211014595952404052.pdf

[4] Korolev, "Open service performance standard (OS PS)," *GLOBAL NAVIGATION SATELLITE SYSTEM GLONASS*, Russia, June 2020. Accessed on: Jul. 9, 2022. [Online]. Available: https://www.glonass-iac.ru/upload/docs/stehos/stehos_en.pdf

[5] "Open service - service definition document," *EUROPEAN GNSS (GALILEO)*, May 2019. Accessed on: Jul. 9, 2022. [Online]. Available: https://galileognss.eu/wp-content/uploads/2020/08/Galileo-OS-SDD_v1.1.pdf

[6] B. Li et al., "LEO enhanced Global Navigation Satellite System (LeGNSS) for real-time precise positioning services," *Advances in Space Research*, vol. 63, no. 1, pp. 73-93, ISSN 0273-1177, 2019.

[7] J. Khalife, M. Neinavaie, and Z. M. Kassas, "Navigation with differential carrier phase measurements from megaconstellation LEO satellites," *IEEE/ION Position, Location and Navigation Symposium (PLANS)*, pp. 1393-1404, 2020.

[8] F. Dovis, L. Lo Presti, and P. Mulassano, "Support infrastructures based on high altitude platforms for navigation satellite systems," *IEEE Wireless Communications*, vol. 12, no. 5, pp. 106-121, Oct. 2005.

[9] O. Kim et al., "Navigation augmentation in urban area by HALE UAV with onboard Pseudolite during multi-purpose missions," *International Journal of Aeronautical and Space Sciences*, 18(3): 545-554, Sep. 2017.

[10] G. Boiero, F. Dovis, P. Mulassano, and M. Mondin, "Increasing the spatial limits of LADGPS using stratospheric platform," *The Journal of Navigation*, vol. 54, no. 2, pp. 255-267, May 2001.

[11] F. Dovis, P. Mulassano, and M. Dumville, "The stratolite concept: Design of a stratospheric pseudo-satellite for Galileo," *Proceedings of ION GPS 2002*, Portland, OR, USA, 2002.

[12] T. Tsujii, M. Harigae, and K. Okano, "A new positioning/navigation system based on Pseudolites installed on high altitude platforms systems (HAPS)," *Proceedings of 24th International Congress of the Aeronautical Sciences*, Yokohama, Japan, 2004.

[13] "Skydel GNSS Simulation Software," *Orolia*, Available: https://www.orolia.com/product/skydel-simulation-engine/.

[14] B. Bidikar, G. S. Rao, and L. Ganesh, "Sagnac effect and SET error based pseudorange modeling for GPS applications," *Procedia Computer Science*, vol. 87, pp. 172-177, 2016.

[15] T. Soler and M. Chin, "On transformation of covariance matrices between local Cartesian coordinate systems and commutative diagrams," *Proceedings of 45th Annual Meeting ASP-ACSM Convention*, Jan. 1985.

[16] A. G. Quinchia, G. Falco, E. Falletti, F. Dovis, and C. Ferrer, "A comparison between different error modeling of MEMS applied to GPS/INS integrated systems," *Sensors*, vol. 13, no. 8, pp. 9549-9588, Jul. 2013.

[17] O. G. Crespillo, M. Joerger, and S. Langel, "Overbounding GNSS/INS integration with uncertain GNSS Gauss-Markov error parameters," *IEEE/ION Position, Location and Navigation Symposium (PLANS)*, Portland, Oregon, pp. 481-489, Apr. 2020.

[18] N. Tongkasem, S. Sopan, J. Budtho, and N. Wongthodsarat, "Calculate user position by using the single point method," *CSSRG Laboratory*, King Mongkut's Institute of Technology Ladkrabang Bangkok, Thailand, Feb. 2019. Accessed on: Jul. 9, 2022. [Online]. Available: https://github.com/cssrg-kmitl/single-positioning-MATLAB

[19] F. Hsieh and M. Rybakowski, "Propagation model for high altitude platform systems based on ray tracing simulation," *Proceedings of 13th European Conference on Antennas and Propagation (EuCAP)*, pp. 1-5, 2019.

[20] S. Alfattani, W. Jaafar, Y. Hmamouche, H. Yanikomeroglu, and A. Yongacoglu, "Link budget analysis for reconfigurable smart surfaces in aerial platforms," *IEEE Open Journal of the Communications Society*, vol. 2, pp. 1980-1995, 2021.

[21] "EVK-M8T user guide," *Ublox*, May 2018. Accessed on: Jul. 9, 2022. [Online]. Available: https://content.u-blox.com/sites/default/files/products/documents/EVK-M8T_UserGuide_%28UBX-14041540%29.pdf

[22] Q. Ren, O. Abbasi, G. K. Kurt, H. Yanikomeroglu, and J. Chen, "Caching and computation offloading in high altitude platform station (HAPS) assisted intelligent transportation systems," *IEEE Transactions on Wireless Communications*, (Early Access), 2022.

[23] C. Ding, J. Wang, H. Zhang, M. Lin, and G. Y. Li, "Joint optimization of transmission and computation resources for satellite and high altitude platform assisted edge computing," *IEEE Transactions on Wireless Communications*, vol. 21, no. 2, pp. 1362-1377, Feb. 2022.

[24] M. S. Alam, G. K. Kurt, H. Yanikomeroglu, P. Zhu, and N. D. Dao, "High altitude platform station based super macro base station constellations", *IEEE Communications Magazine*, vol. 59, no. 1, pp. 103-109, Jan. 2021.

[25] A. Yastrebova, M. Höyhtyä, S. Boumard, E. S. Lohan, and A. Ometov, "Positioning in the Arctic region: State-of-the-art and future perspectives," *IEEE Access*, vol. 9, pp. 53964-53978, Mar. 2021.